\documentclass[twocolumn,a4paper]{revtex4-1}
\usepackage{hyperref}
\usepackage{graphicx}
\usepackage{amsmath}
\usepackage{xcolor}
\usepackage{float}
\usepackage{amsfonts}
\usepackage{epsfig}
\usepackage{graphics}
\usepackage{color}
\usepackage{csquotes}

\newcommand{\beq}{\begin{equation}}
\newcommand{\eeq}{\end{equation}}
\newcommand{\bqa}{\begin{eqnarray}}
\newcommand{\eqa}{\end{eqnarray}}

\begin{document}
\title{Dissociation of heavy quarkonia at finite $\mu$}
\author{Siddhartha Solanki$^{a}$} 
\author{Vineet Kumar Agotiya$^{a}$}
\email{agotiya81@gmail.com}
\affiliation{$^a$Department of Physics, Central University of Jharkhand, Ranchi, India, 835222}

\begin{abstract}
We studied the properties of the heavy quarkonia in the presence of finite quark-chemical potential for different number of flavors by using the quasi particle approach. The effect of the finite quark-chemical potential has been incorporated through the quasi-particle Debye mass to examine the binding energies of the quarkonium states. From the imaginary part of the potential we have calculated the thermal width of the ground state of the quarkonia and found that the thermal width increases with finite quark-chemical potential. The dissociation temperature $(T_{D})$ of the $J/\psi$ and $\Upsilon$ have been calculated in the presence of finite quark-chemical potential for different flavors (i.e., $N_{f}$=1, 2 and 3). The effect of the finite quark-chemical potential on the mass spectra of the quarkonium states has been also studied.\\

\end{abstract}

\maketitle

\section{Introduction}
The ongoing heavy ion collision experiment at the relativistic heavy ion collider (RHIC) accelerator in BNL, USA and the Large Hadron Collider (LHC) collider at CERN, Switzerland has their great importance for exploring the phase diagram of the Quantum Chromo-dynamics (QCD) after the discovery of the Quark Gluon Plasma (QGP), the fourth state of matter. The study of the fundamental forces between quark and gluon is essential for the understanding of QCD and it has been expected that at varying temperature, finite quark-chemical potential scales (low or high) and different phases contribute in the T-$\mu$ plane. For instance, at small or vanishing temperature quarks and gluons are limited by the strong force while at high temperature asymptotic freedom suggests a somewhat different QCD medium which consists of rather weakly coupled deconfined quarks and gluons, the so-called QGP. The anomalous suppression of the $J/\psi$ production in heavy-ion collisions which has been observed experimentally~\cite{M.C.Abreu} in the depletion of the dilepton multiplicity in the region of invariant mass which corresponds to the $J/\psi$ meson was proposed long back as a possibly unmistakable sign of the beginning of deconfinement. Matsui and Satz~\cite{T.Matsui} argued that charmonium states generated before the formation of a thermalized QGP would tend to melt in their way through the deconfined medium, because the binding coulomb potential is screened by the large number of color charges, thereby producing an anomalous drop in the $J/\psi$ yields. The pair develops into the physical resonance during formation time and passes through the plasma and the hadronic matter before they leave the interacting system to decay into a dilepton to be detected.\\
This long 'trek' within the interacting system is fairly dangerous for the $J/\psi$. Even before the resonance occurs, it may be absorbed by the nucleons streaming past it~\cite{C.Gerschel}. By the time the quarkonium resonance is formed, the screening of the color forces in the plasma may be sufficient to inhibit a binding of the charmonium~\cite{T.Matsui} or an energetic gluon~\cite{X.M.Xu} or a comoving hadron could dissociate the resonance. Quarkonia at finite temperature is an important tool for the studying QGP formation in heavy-ion collisions. Many effort have been devoted to determine the $T_D$ of quarkonium state in the deconfined medium, using either lattice calculations of $Q\bar{Q}$ spectral function or non-relativistic calculations based upon some effective screened potential. Lattice studies are directly based on QCD and these studies answer most of the questions that arises while studying the QCD phase diagram. However, in lattice studies the spectral function must be extracted using rather limited sets of data from the Euclidean (imaginary time) correlators, which are directly based on the lattice. This along with the intrinsic technical difficulties of lattice calculations, limit the reliability of the results obtained so far, and also their scope is essentially limited to the mass of the ground states in each quarkonium channel. Potential models, on the other hand, provide a simple and intuitive framework to study the properties of quarkonium at finite temperature, from which quantities can be calculated that are beyond the current possibilities for lattice studies. Umeda and Alberico~\cite{T.Umeda, W.M.Alberico(2008)} have shown that the lattice computations of mesons correlators at finite temperature contain a constant contribution, due to the presence of zero modes in the spectral functions. The problem of dissociation of bound states in a hot QCD medium is of great importance in heavy-ion collisions as it provides evidence for the creation of the QGP~\cite{Leitch}. The physical understanding of the quarkonium dissociation within a deconfined medium has undergone several refinements in the last couple of years~\cite{Laine:2008cf, BKP:2000, BKP:2001, BKP:2002, BKP:2004}. As the heavy quark and anti-quark in a quarkonia state are bound together by almost static (off-shell) gluons, therefore, the issue of their dissociation boils down to how the gluon self-energy behaves at high temperature. It has been noticed that the gluon self-energy has both real and imaginary parts~\cite{laine}. Note that the real part lead to the Debye screening while the imaginary part leads to Landau damping and give rise the thermal width to the quarkonia.\\
It indeed provides a useful way to examine quarkonium binding energies, quarkonium wave functions, reaction rates, transition rates, and decay widths. It further allows the extrapolation to the region of high temperatures by expressing screening effects reflecting on the temperature dependence of the potential and finite quark-chemical potential. The effects of dynamics of quarks on the stability of quarkonia can be studied by using finite quark-chemical potential extracted from thermodynamic quantities that are computed in full QCD. At high temperatures, the deconfined phase of QCD exhibits screening of static color-electric fields~\cite{E.V.Shuryak, GPY}; it is, therefore, expected that the screening will lead to the dissociation of quarkonium states. After the success at zero temperature while predicting hadronic mass spectra, potential model descriptions have been also applied to understand quarkonium properties at finite temperature and finite quark-chemical potential. It is well known that the production of $J/\psi$ and $\Upsilon$ mesons in hadronic reactions occur in part via the production of higher excited $c\bar{c}$ (or $b\bar{b}$) states and they decay into the respective ground state. Since the lifetime of different quarkonium state is much larger than the typical lifetime of the medium produced in nucleus-nucleus collisions; their decay occurs almost completely outside the produced medium~\cite{lain, he1}. The produced medium can be probed not only by the ground state quarkonium but also by different excited quarkonium states. So, the potential model in this context could be helpful in predicting the binding energies of various quarkonia state by setting up and solving appropriate Schrodinger equation in the hot QCD medium. Recently a novel work was carried out by  Bo Tong and Baoyi Chen~\cite{B.Tong} in which Schrodinger equation have been solved  to study the evaluation of charmonium wave package at finite baryonic chemical potential. The chemical potential correction is included in the Debye mass which is employed in both real and imaginary part of the potential for  studying the properties of quarkonia in our present work are similar to the study~\cite{B.Tong}. Thereafter the dissociation of ground states and excited states of quarkonia has been studied.\\
The present work is different from the work~\cite{U.Kakade} in which the authors have studied the effect of baryonic chemical potential on the Quarkonium states using leading order Debye mass and they have  calculated the dissociation chemical potential (is in the unit of $\mu_{c}$) for the J/$\psi$ state at different temperature. In the current work, we have started with the modified form of the potential~\cite{V.Agotiya:2009} in which the effect of the finite quark-chemical potential has been incorporated through the quasi-particle Debye mass. The energy eigen values of the ground states of the charmonium and bottomonium for different flavors has been obtained by solving the Schrodinger equation. Further we have studied the effect of the finite quark-chemical potential on the thermal width and calculated the dissociation temperature (is in the unit of $T_{c}$) for the quarkonium states at different range of quark-chemical potential.\\
The present manuscript is organized in the following manner: a brief description  about the real and the imaginary part of the heavy quark potential is given in section-II. Whereas in section-III  finite quark-chemical potential is introduced into the  quasi-particle Debye mass. In Section-IV, we examine the  binding energy and $T_D$ of various quarkonium states at different values of flavor and the finite quark-chemical potential. The effect of the finite quark chemical  potential on the mass spectra of the quarkonium states is evaluated in the section-V. In section-VI we discuss the result of present work. Finally, in section-VII, we conclude with the future prospects of the present work.

\section{Heavy quark complex potential}

\subsection{Real Part of the potential}
Quantitative understanding of the bound state properties needs the exact potential at finite-temperature which should be derived directly from QCD, like the Cornell potential at zero temperature has been derived from pNRQCD from the zeroth-order matching coefficient. Such derivations at finite temperature for weakly-coupled plasma have been recently come up in the literature~\cite{Brambilla05,Brambilla08} but they are plagued by the existence of temperature-driven hard as well as soft scales, $T$, $gT$, $g^2 T$, respectively. Due to these difficulties in finite temperature extension in effective field theories, the lattice-based potentials become popular.\\
\begin{figure*}   
    \includegraphics[height=6.5cm,width=6.5cm]{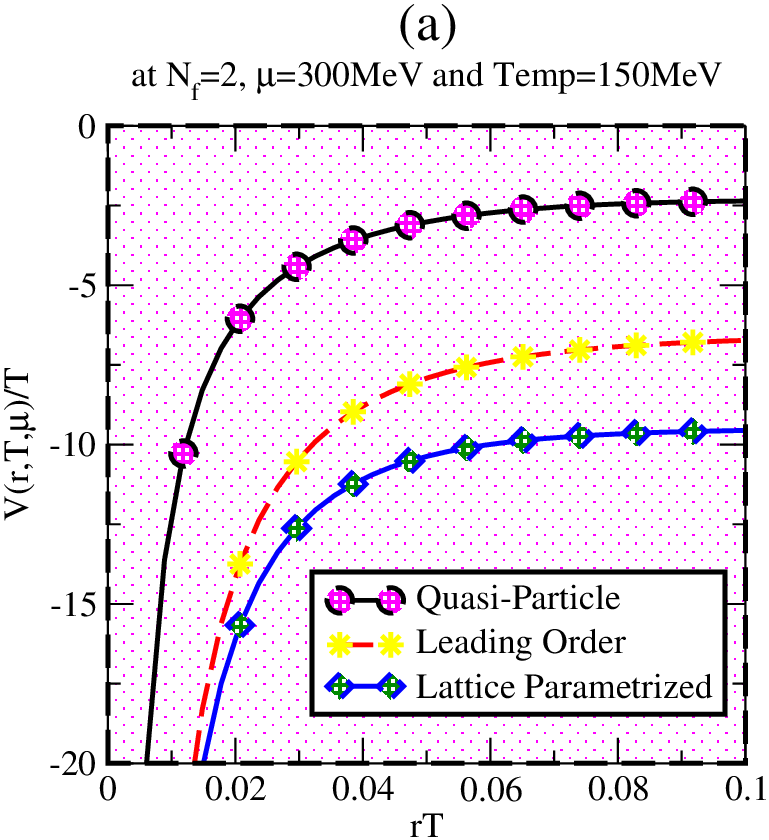} 
    \hspace{3cm}
    \includegraphics[height=6.5cm,width=6.5cm]{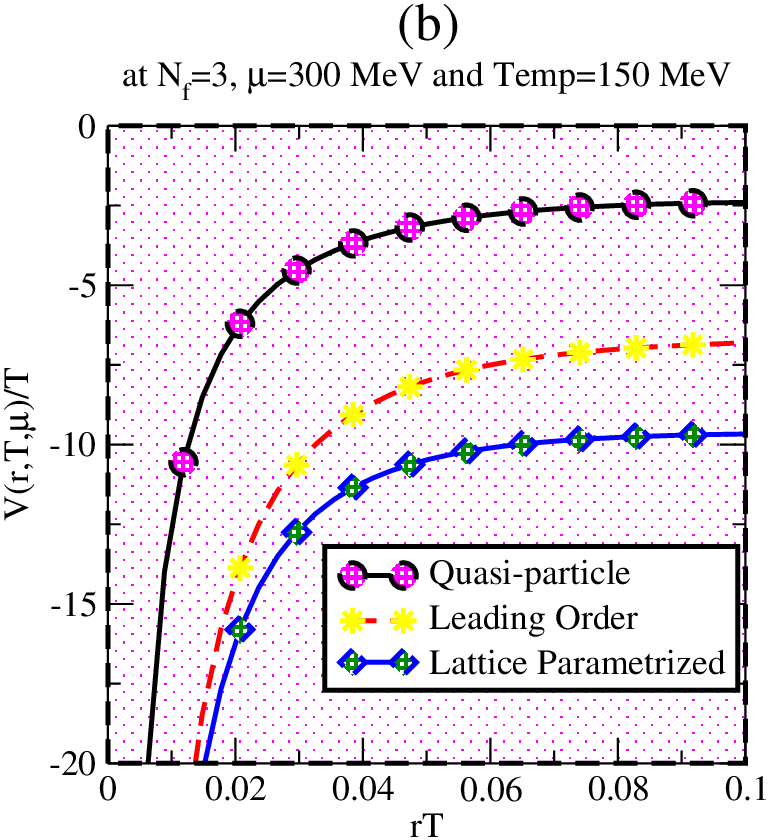}
    \vspace{5mm}
\caption{The variation of potential with the distance 'r' for the quasi-particle, leading order and lattice parameterized Debye mass at $N_f$= 2 and 3 has been shown.} 
\label{fig.1} 
\vspace{5mm} 
\end{figure*}
However, neither the free energy nor the internal energy can be directly used as the potential. What kind of screened potentials should be used in the Schrodinger equation which well describes  the bound states at finite temperature is still an open question. The potential model based phenomenology as well as the lattice QCD approaches infers that the quark-antiquark interaction potential in presence of medium plays a crucial role in understanding the nature of quark-antiquark bound state in the hot QCD/QGP medium. The potential employed is commonly screening columns (Yukawa form)~\cite{Brambilla05,L.Kluberg}. In case of finite temperature QCD we employ the ansatz that the medium modification enter in the Fourier transform of heavy quark potential V(k) as~\cite{V.Agotiya:2009}. Next, to modify the potential, we first need to calculate the dielectric permittivity, which is obtained from the self-energy using finite temperature QCD. It is important to note that the perturbative theory at $T>0$ suffers from the infrared singularities and gauge-dependent results because the perturbative expansion is incomplete at $T>0$. There are infinitely many higher order diagrams with more and more loops that can contribute to lower order in the coupling constant.
\begin{equation}
\label{eq1}
\tilde{V(k)}=\frac{V(k)}{\varepsilon(k)}
\end{equation}
Where, $\varepsilon(k)$ is dielectric permitivity which is obtain from the static limit of the longitudinal part of the gluon self energy~\cite{R.A.Schneider, H.A.Weldon}.
\begin{equation}
\label{eq2}
\varepsilon(k)=\left(1+\frac{\pi_L(0,k,T)}{k^2}\right)\equiv\left(1+\frac{m^2_D(T,\mu)}{k^2}\right)
\end{equation}
V(k) is the Fourier transform of the Cornell potential given as,
\begin{equation}
\label{eq3}
V(k)=-\sqrt{\frac{2}{\pi}}\frac{\alpha}{k^2}\\-\frac{4\sigma}{\sqrt{2\pi}k^4}
\end{equation}
Substituting the value of Eq.(\ref{eq2}) and Eq.(\ref{eq3}) in equation Eq.(\ref{eq1}), and solving this by using inverse Fourier transform, we get the medium modified potential depending upon 'r', temperature, and quark chemical potential~\cite{V.Chandra:2007, R.A.Schneider, A.Ranjan} and is given by:
\begin{multline}
\label{eq4}
V(r,T,\mu)=\left(\frac{2\sigma}{m^2_D(T,\mu)}-\alpha\right)\frac{exp(-m_D(T,\mu)r)}{r}\\-\frac{2\sigma}{m^2_D(T,\mu)r}+\frac{2\sigma}{m_D(T,\mu)}-\alpha m_D(T,\mu)
\end{multline}
Where $\alpha$ is the coupling constant and $\sigma$=$0.184$ $GeV^{2}$ is the string coefficient.\\

\subsection{Imaginary Part of the potential}
Heavy quark complex potential contain both the real and the imaginary part which were obtained by using the gluon self energy which in turn responsible for both the Debye screening and the Landau damping respectively. However, Debye screening is obtained by using both the retard and self-energy propagator whereas the static limit of the symmetric self-energy has been used for calculating the imaginary part of the potential. The imaginary part of the potential has its importance while studying the threshold enhancement of the bound state or the thermal width of the charmonium and bottomonium resonances. This thermal width in spectral function is further used to determine the $T_{D}$ of the quarkonium resonances. From the previous studies~\cite{AgnesMocsy, M.LaineJHEP2007} is it clear that, the dissociation of the quarkonium states takes place when thermal width becomes equal to the twice  of the binding energy.
   \begin{figure*} 
\vspace{2cm}  
    \includegraphics[height=6.5cm,width=6.5cm]{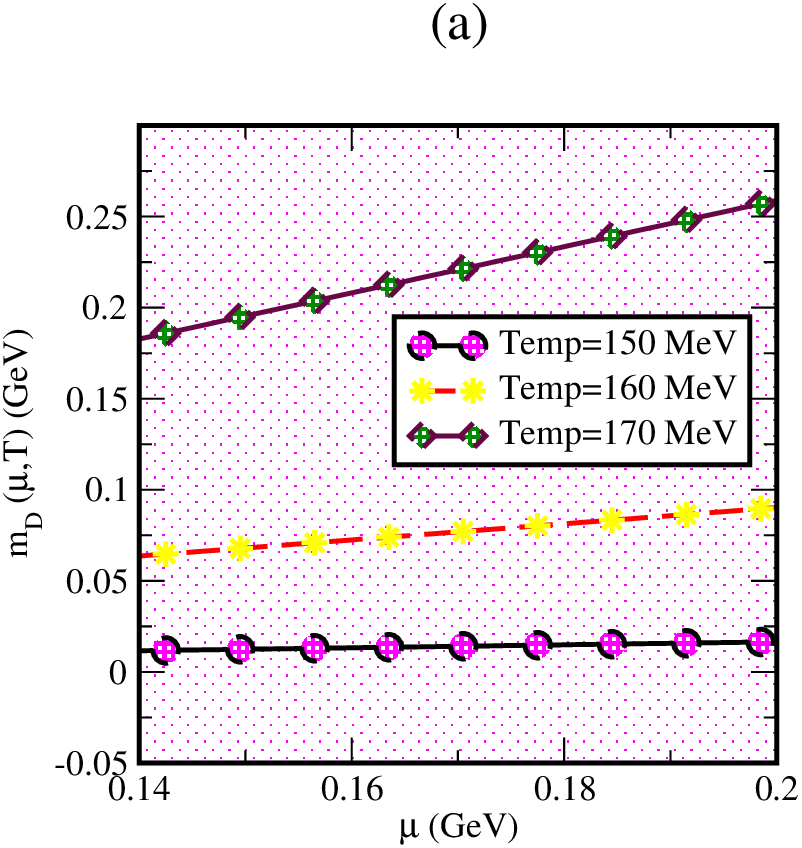}
    \hspace{3cm}
    \includegraphics[height=6.5cm,width=6.5cm]{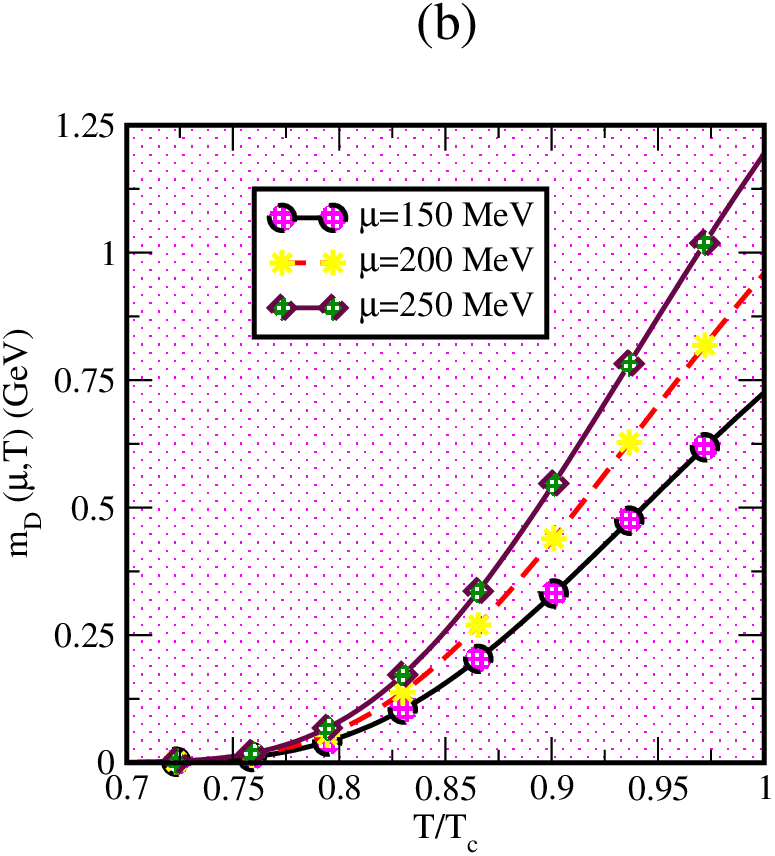}
    \vspace{5mm}
    \caption{Shows the variation of quasi-particle Debye mass with finite quark-chemical potential at different values of temperature 2 (a) and with temperature at different values of finite quark-chemical potential 2 (b) respectively.}
\label{fig.2}
\vspace{2.5cm}   
\end{figure*}
\begin{figure*}  
\vspace{2cm} 
    \includegraphics[height=6.5cm,width=6.5cm]{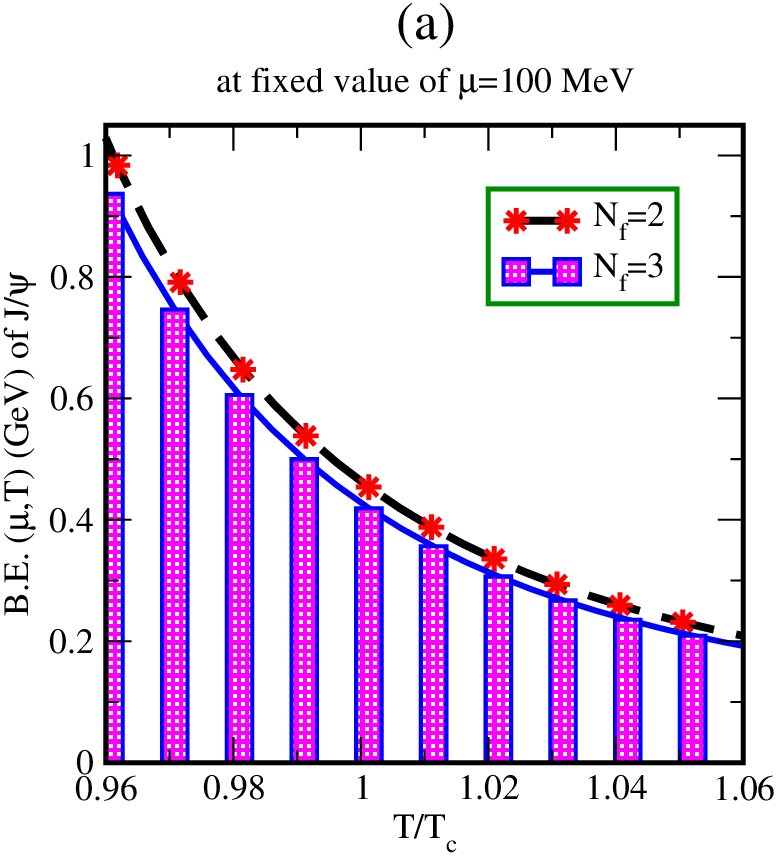}
    \hspace{3cm}
    \includegraphics[height=6.5cm,width=6.5cm]{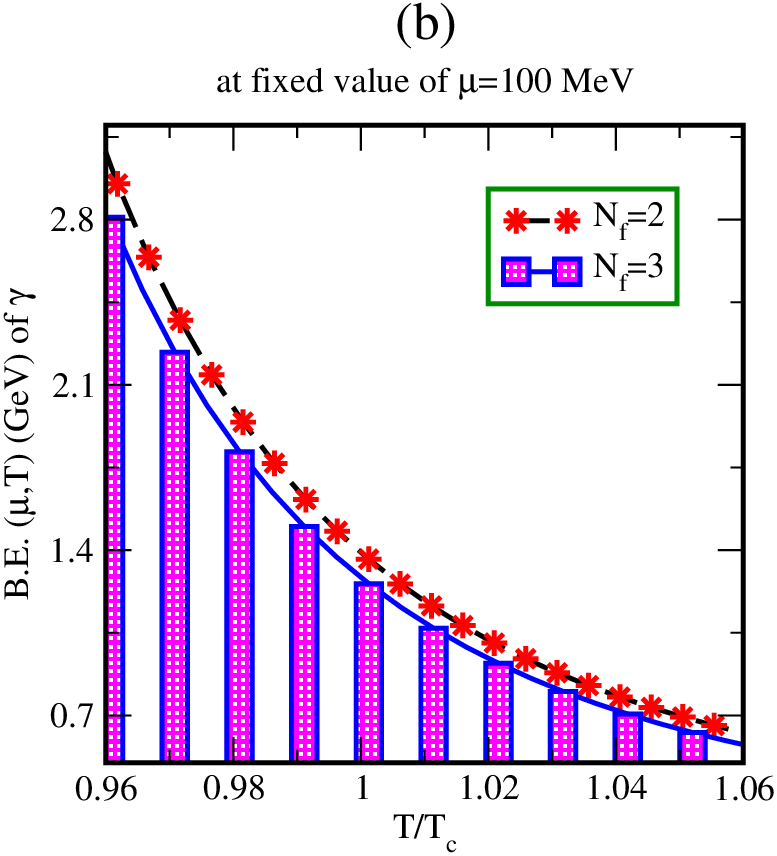}
    \vspace{5mm}
\caption{Shows the variation of binding energy of $J/\psi$  and $\Upsilon$ with the temperature in figures 3 (a) and 3 (b) for different flavors at $N_{f}$ = 2 and 3.}
\label{fig.3} 
\vspace{2cm}
\end{figure*}
The symmetric propagator for the imaginary part is given below:
\begin{equation}
\label{eq5}
ImD_F^{00}(0,p)=\frac{-2\pi{T} {m_D^2}}{p\left(p^2+m_D^2\right)^2}
\end{equation}
Eq.\ref{eq5} gives the imaginary part of the dielectric function as;
\begin{equation}
\label{eq6}
\epsilon^{-1}(p)=-\pi{T}{m_D^2}\ \frac{p^2}{p\left(p^2+m_D^2\right)^2}
\end{equation}
Thus, we obtained the imaginary part of the potential using the following equation:
\begin{equation}
\label{eq7}
V(r,T,\mu)=\int{\frac{d^3k}{\left(2\pi\right)^\frac{3}{2}}\left(e^{i.pr}-1\right)\frac{V(p)}{\epsilon(p)}}
\end{equation}
This implies
\begin{multline}
\label{eq8}
ImV\left(r,T,\mu\right)=\int{\frac{d^3k}{\left(2\pi\right)^\frac{3}{2}}\left(e^{i.pr}-1\right)\left(-\sqrt{\frac{2}{\pi}}\frac{\alpha}{p^2}-\frac{4\sigma}{\sqrt{\pi}{p^4}}\right)}\\\times p^{2}\left( {\frac{-\pi^{2}{T} {m_D^2}}{p\left(p^2+m_D^2\right)^2}} \right)
\end{multline}
After performing the integration of the above Eq.\ref{eq8}, the contribution due to coulomb and the string term becomes:
\begin{multline}
\label{eq9}
ImV\left(r,T,\mu\right)= -2{\alpha}{T}\int_{0}^{\infty}\\{\frac{dz}{{(z^2+1)}^2}\left(1-\frac{\sin{z}}{z\hat{r}}\right)+\frac{4\sigma{T}}{m_D^2}\int_{0}^{\infty}{\frac{dz}{{(z^2+1)}^2}\left(1-\frac{\sin{z\hat{r}}}{z\hat{r}}\right)}}
\end{multline}
Where z = $\frac{P}{m_D}$. This can be further simplified  as:
\begin{equation}
\label{eq10}
ImV\left(r,T,\mu\right)=\-{\alpha}{T}\phi_0\left(\hat{r}\right)+\frac{2\sigma{T}}{m_D^2}\psi_0\left(\hat{r}\right)
\end{equation}
Where  $\phi_0\left(\hat{r}\right)=-{\alpha}{T}\left(\frac{{\hat{r}}^2}{9}\left(-4+3\gamma_E+ 3log(\hat{r})\right)\right)$\
~\\
 and 
 ~\\
 $\psi_0\left(\hat{r}\right)=-\frac{{\hat{r}}^2}{6}\ +\left(\frac{-107+60\gamma_E+60log(\hat{r})}{3600}\right){\hat{r}}^4+O({\hat{r}}^5)$
~\\
For the limit $\hat{r}<<1$, we have
\begin{equation}
\label{eq11}
ImV\left(r,T,\mu\right)=-T\left(\frac{\alpha{\hat{r}}^2}{3}+\frac{\sigma{\hat{r}}^4}{30m_D^2}\right)log\left(\frac{1}{\hat{r}}\right)
\end{equation}
In the small-distance limit, the imaginary part of the potential can be considered as a perturbation to the vacuum potential~\cite{Thakur}, which provides an estimate for the thermal width for a particular resonance state. The medium potential, at high temperature, has a long range Coulombic tail that dominates over all the other terms. Owing to this fact, one can choose the $\Psi(r)$ as the Coulombic wave function. The Coulombic wave function for the ground state [1s, corresponding to n=1 $J/\psi$ and $\Upsilon$)]  can be written:
$\psi_{1s}(r)=\frac{1}{\sqrt{\pi a_{o}^{3}}}e^{-\frac{r}{a_{o}}}$,where $a_{o}=\frac{2}{\alpha m_{Q}}$ is the Bohr's radius of the respective quarkonium states.
Now with the imaginary potential defined Eq.\ref{eq11} and the coulombic wave function for the ground states defined above, we calculate the thermal width $\Gamma$ of the resonance state (1S). This can be done by folding with unperturbed (1S) Coulomb wave function and is given by~\cite{prd2016_vin},
\begin{equation}
\label{eq12}
\Gamma=\left(\frac{4T}{\alpha{m_Q^2}}+\frac{12\sigma{T}}{\alpha^2m_Q^4}\right)m_D^2log\frac{\alpha{m_Q}}{2m_D}
\end{equation}
We have now both the real and the imaginary of the potential depending upon quark chemical potential, so by using the criteria discussed earlier, we calculate the dissociation temperature for the quarkonium states. This can be done by exploiting the twice the binding energy along with the thermal width for the particular state.
\begin{figure*}   
    \includegraphics[height=6.5cm,width=6.5cm]{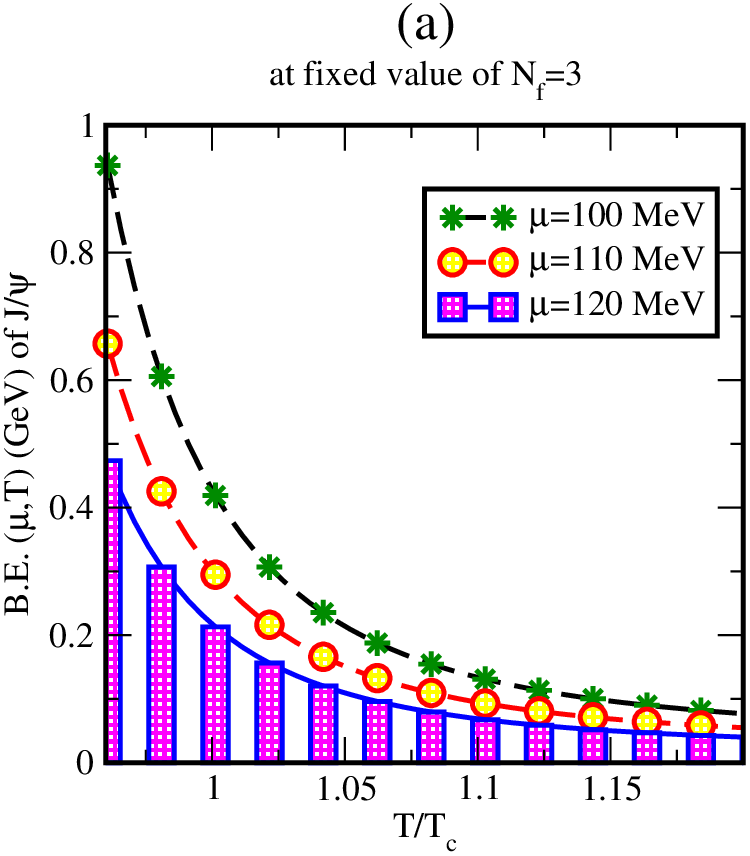}
    \hspace{3cm}
    \includegraphics[height=6.5cm,width=6.5cm]{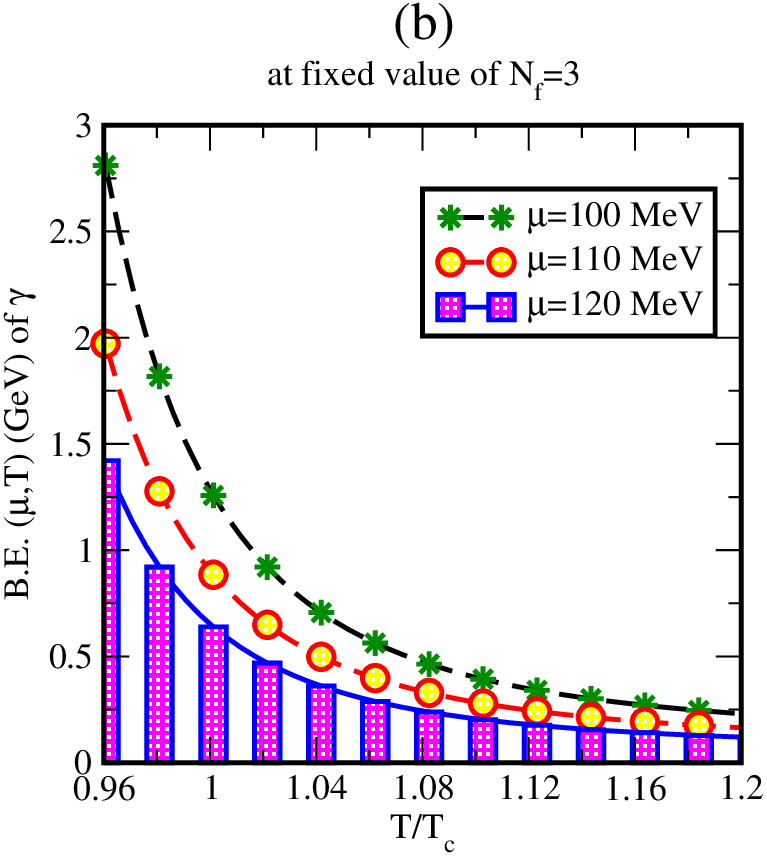}
    \vspace{5mm}
\caption{Shows the variation of binding energy of $J/\psi$  and$\Upsilon$ with the temperature in the  figures 4 (a) and 4 (b) for different values of the finite quark-chemical potential at $N_f$= 3.}
\label{fig.4}
\vspace{2cm} 
\end{figure*}
\begin{figure*} 
\vspace{2cm} 
    \includegraphics[height=6.5cm,width=6.5cm]{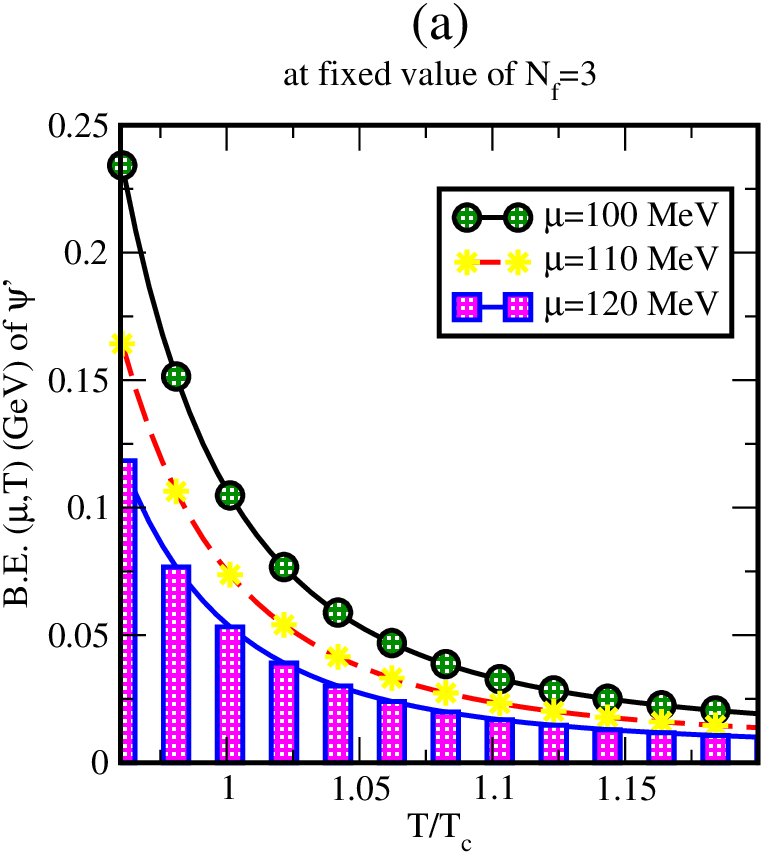}
    \hspace{3cm}
    \includegraphics[height=6.5cm,width=6.5cm]{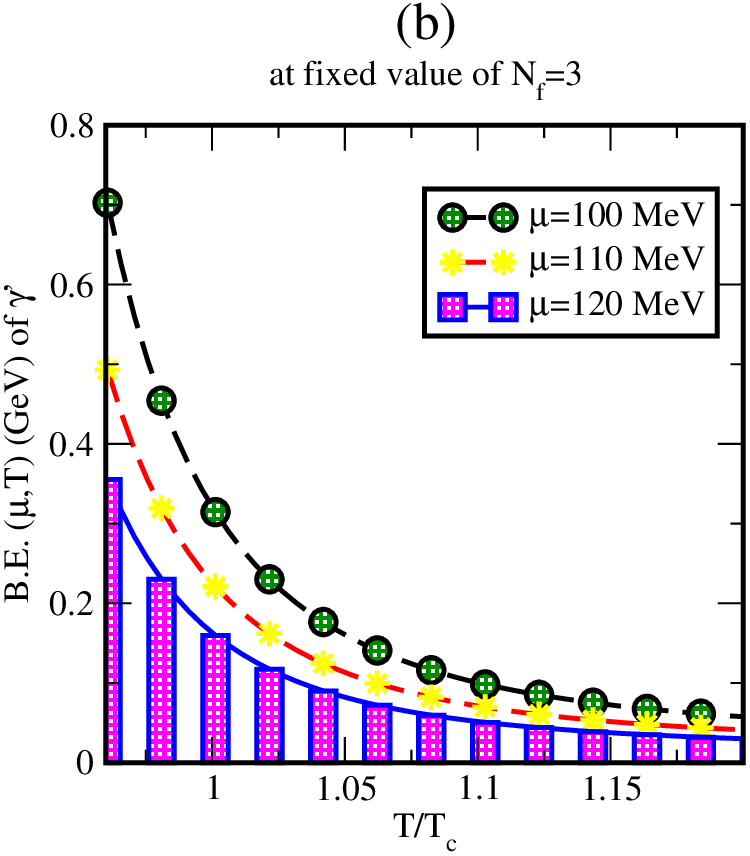}
    \vspace{5mm}
\caption {Shows the variation of Binding energy of $\psi^{\prime}$ and $ \Upsilon^{\prime}$ with the temperature in the figures 5 (a) and 5 (b) for different values of the finite quark-chemical potential at $N_f$= 3.}
\label{fig.5}
\vspace{5mm}
\end{figure*}

\section{Quasi-particle Debye mass in the presence of finite quark chemical  potential}
The perturbative nature of the leading order Debye mass in QCD coupling at high temperature is known for a long time~\cite{A.Rebhan}. 
In QCD the Debye mass is non-perturbative and gauge independent defined in~\cite{E.Braaten}. Debye mass is also calculated for the two polyakov loops by Braaten and Nicto at high temperature~\cite{Y.Burnier}. The basic definition of the Debye mass itself creates a difficulty because of the gauge variant nature of electric correlators in~\cite{K.Kajantie}.\\
To overcome this problem many approaches have been purposed so far~\cite{K.Kajantie, Anbazavov, S.Nadkarni}. Keeping in view of all the interaction between the quasiparticle because of the quasi-parton, a number of attempts have been made so far such as, effective mass model~\cite{V.Goloviznin, A.Peshier}, the effective mass  with Polyakov loop~\cite{M.D.Elia}, model based on PNJL and NJL~\cite{A.Dumitru}, effective fugacity model~\cite{V.Chandra:2007, V.Chandra:2009}. Quasi-particle model is important to describe the non-ideal behavior of the QGP and the masses arise because of the surrounding matter in the medium around the parton and this quasi-parton acquired the same quantum number as the real particle i.e quarks and gluon~\cite{P.K.Srivastava}. Here we use quasi-particle EoS~\cite{M.Cheng}. The quasi-particle Debye mass ${m_D}$ in terms of the temperature and finite quark-chemical potential for full QCD case is given as,
\begin{figure*} 
\vspace{2cm} 
    \includegraphics[height=6.5cm,width=6.5cm]{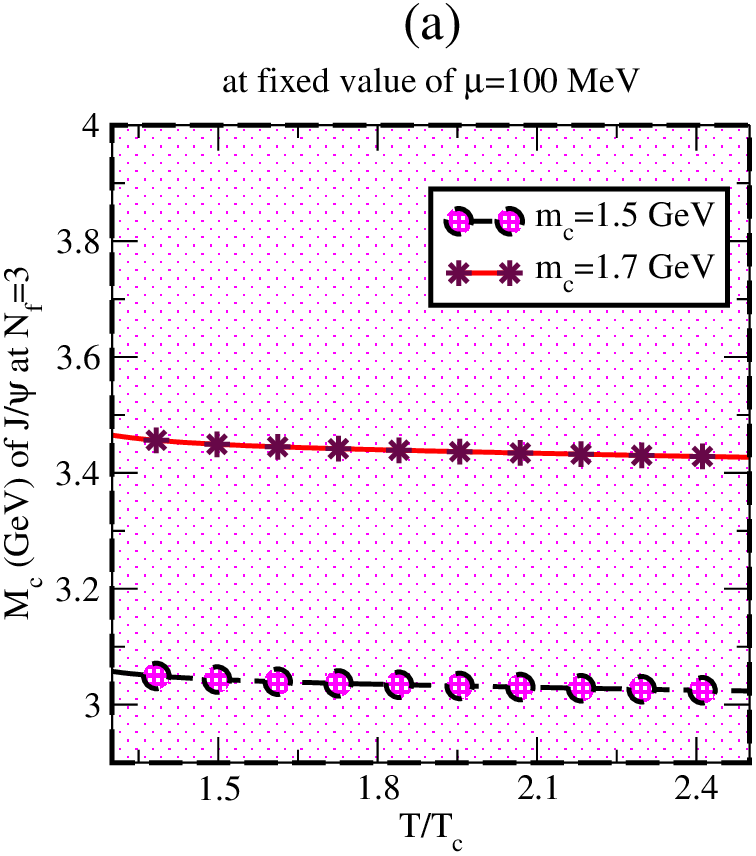}
    \hspace{3cm}
    \includegraphics[height=6.5cm,width=6.5cm]{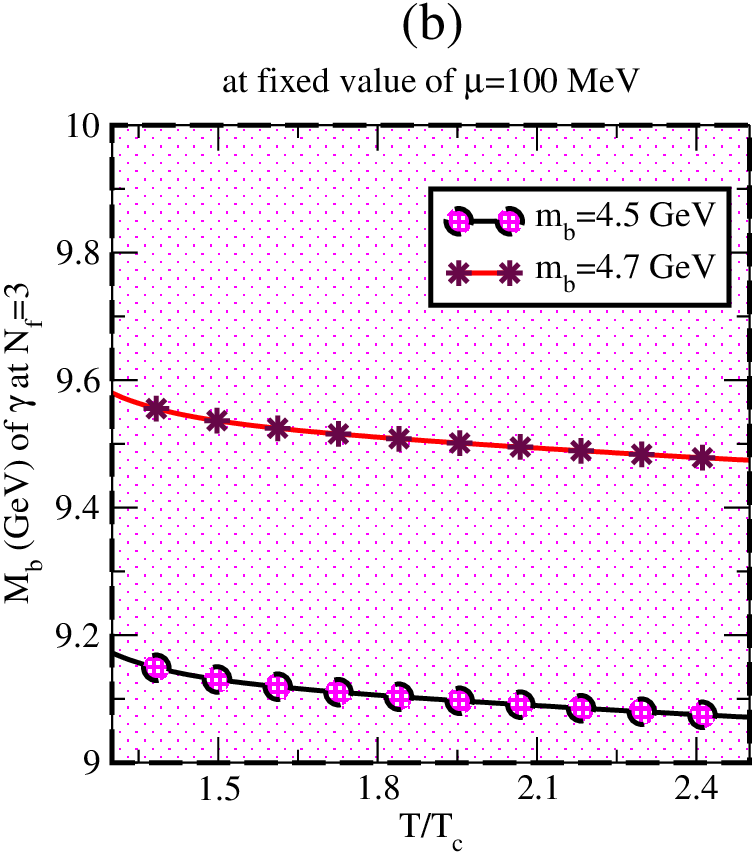}
    \vspace{5mm}
\caption{Mass spectra of $J/\psi$ and $\Upsilon$ with the temperature in the figures 6 (a) and 6 (b) for different masses has been shown.}
\label{fig.6}  
\vspace{5mm}
\end{figure*}
\begin{eqnarray}
\label{eq13}
m^2_D\left(T,\mu\right) &=& g^2(T) T^2 \bigg[
\bigg(\frac{N_c}{3}\times\frac{6 PolyLog[2,z_g]}{\pi^2}\bigg)\nonumber\\&&
+{\bigg(\frac{\hat{N_f}}{6}\times\frac{-12 PolyLog[2,-z_q]}{\pi^2}\bigg)\bigg]}
\end{eqnarray}
and the value of $\hat{N_f}$ is,
\begin{eqnarray}
\label{eq14}
\hat{N_f} &=& \bigg[N_f +\frac{3}{\pi^2}\bigg(\frac{\mu^2}{T^2}\bigg)\bigg]
\end{eqnarray}
and
\begin{eqnarray}
\label{eq15}
\mu &=& \frac{\mu_b}{3}
\end{eqnarray}
Where $(\mu)$ defined the finite quark-chemical potential and $(\mu_b)$ is finite baryonic chemical potential and can be found in the reference~\cite{U.Kakade,Solanki,Solanki2022}. After introducing the value of $\hat{N_f}$ in the Eq.(\ref{eq13}), we get the full expression of quasi-particle Debye mass in terms of temperature and finite quark-chemical potential as, 
\begin{equation}
\label{eq16}
m^2_D\left(T,\mu\right)=g^2(T) T^2 \left\{ \bigg( \frac{N_c}{3} Q^2_g\bigg)+\bigg[\frac{N_f}{6}+\frac{1}{2\pi^2}\bigg(\frac{\mu^2}{T^2}\bigg)\bigg]Q^2_q\right \}
\end{equation}
Where, $Q_g$ and $Q_q$ are the effective charges given by the equations:
\begin{equation}
\label{eq17}
z_{g,q}= a_{q,g}\exp\bigg(-\frac{b_{g,q}}{x^2}-\frac{c_{g,q}}{x^4}-\frac{d_{g,q}}{x^6}\bigg).
\end{equation}
Here $x=T/T_c$ and $a$, $b$ and $c$ and $d$ are fitting parameters, for equation of state in the quasi-particle description~\cite{V.Chandra:2007, V.Chandra:2009} respectively.
\begin{eqnarray}
\label{eq18}
 Q^2_g&=&\frac{6 PolyLog[2,z_g]}{\pi^2}\nonumber\\
 Q^2_q&=&\frac{-12 PolyLog[2,-z_q]}{\pi^2}
\end{eqnarray}
\begin{figure*}   
    \includegraphics[height=6.5cm,width=6.5cm]{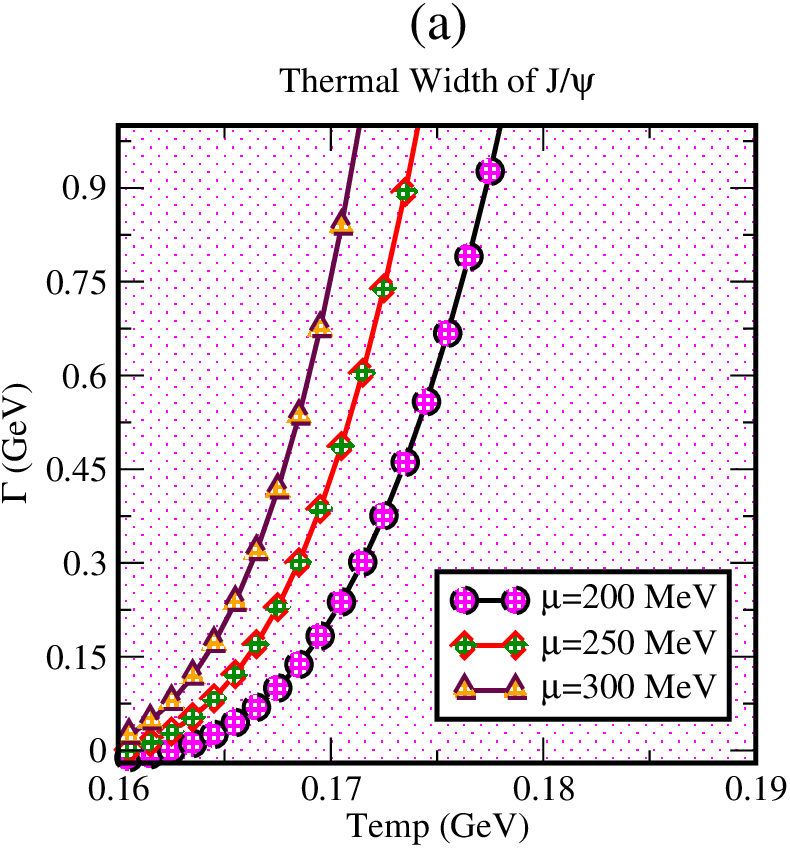}
    \hspace{3cm}
    \includegraphics[height=6.5cm,width=6.5cm]{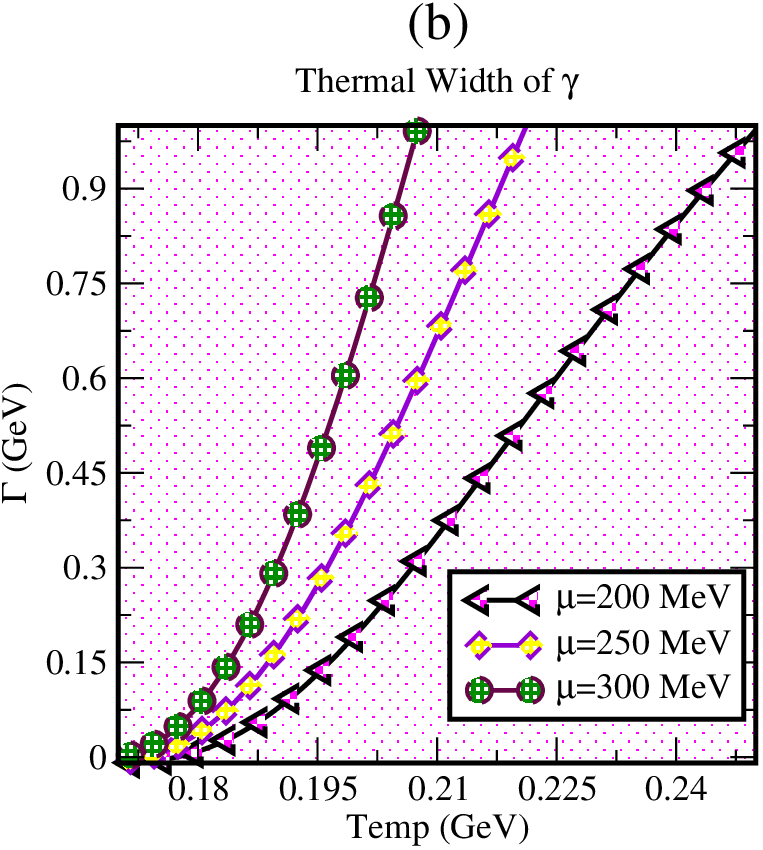}
    \vspace{5mm}
\caption{Shows variation of the $J/\psi$ and $\Upsilon$ thermal width $\Gamma$ with the temperature in the figures 7 (a) and 7 (b) for different finite quark-chemical potential.}
\label{fig.7} 
\vspace{2cm}
\end{figure*}
\begin{figure*}  
\vspace{2cm} 
    \includegraphics[height=6.5cm,width=7cm]{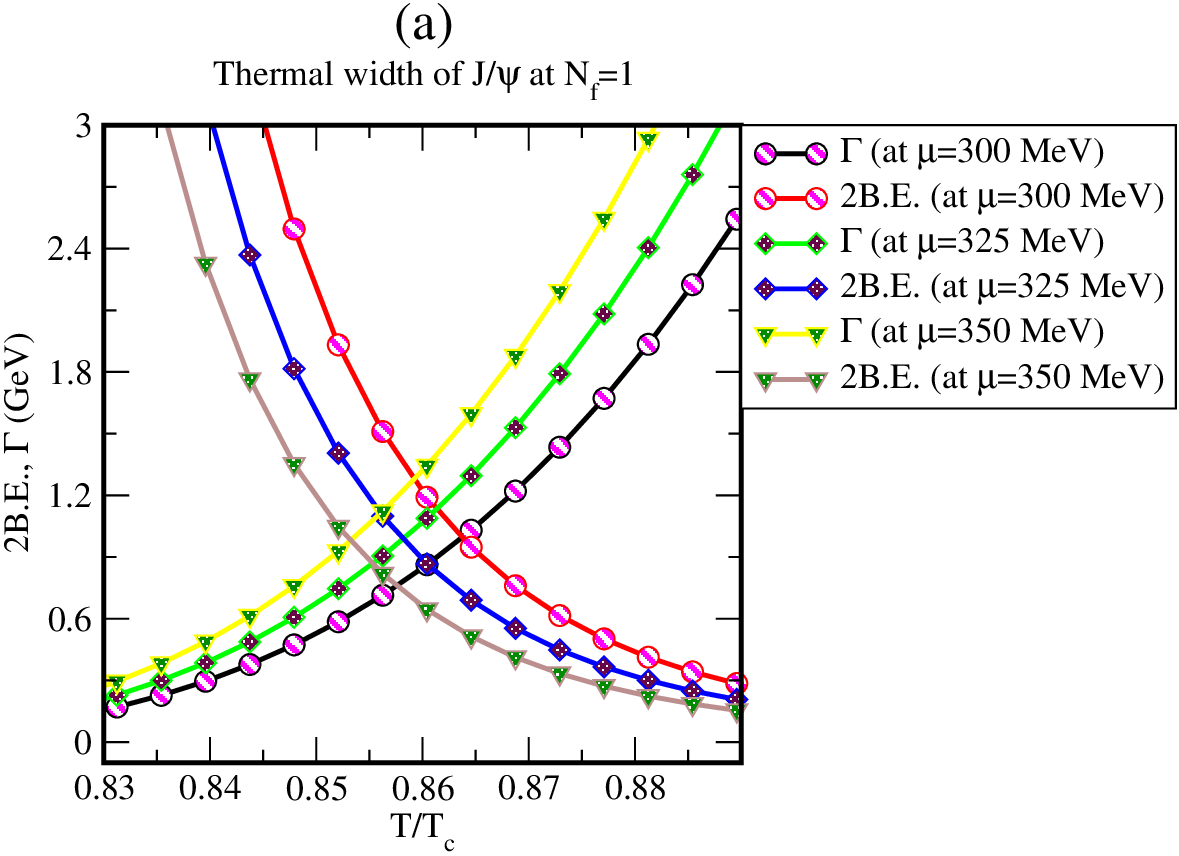}
    \hspace{1cm}
    \includegraphics[height=6.5cm,width=7cm]{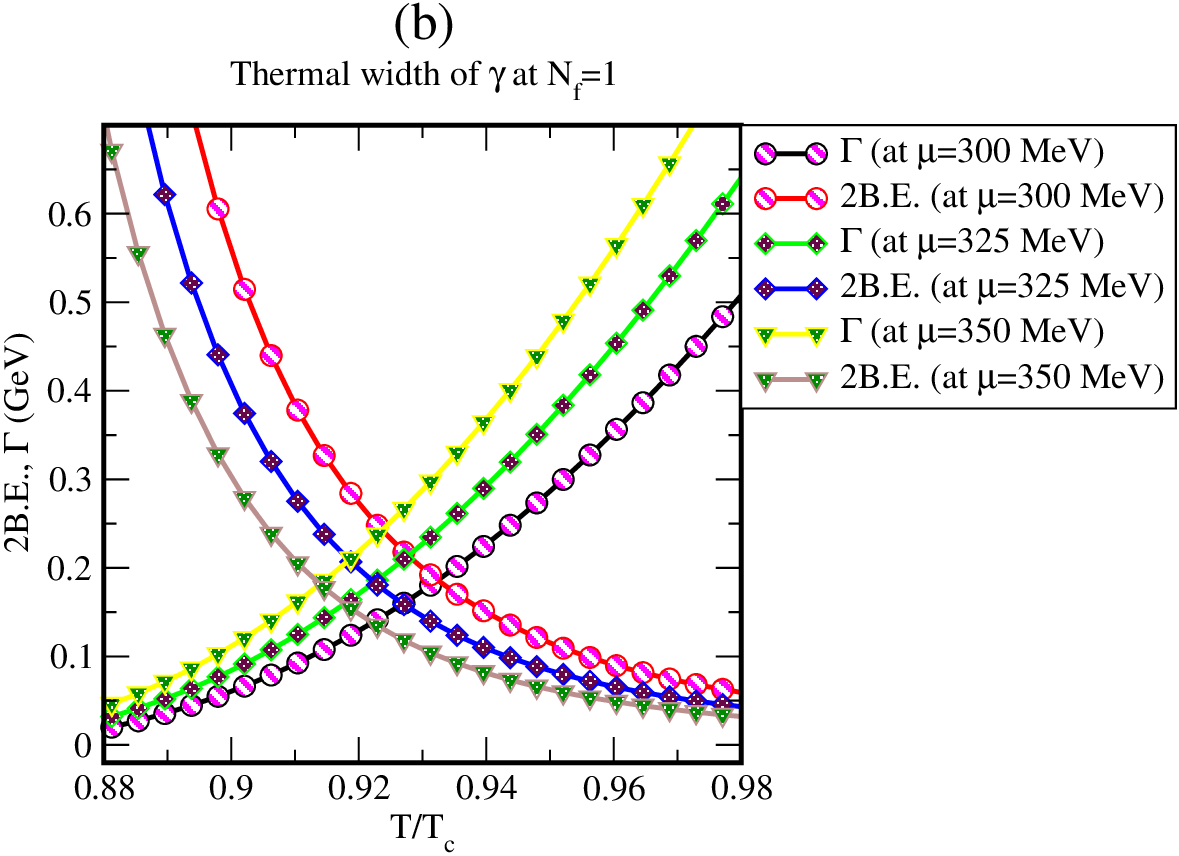} 
    \vspace{5mm} 
\caption{Shows the real and imaginary binding energies of the $J/\psi$ and $\Upsilon$ for different finite quark-chemical potential at $N_f$= 1 in the figures 8 (a) and 8 (b).}
\label{fig.8} 
\vspace{5mm}  
\end{figure*}
Here, $g(T)$ is the QCD running coupling constant, $N_c$=3 ($SU(3)$) and $N_f$ is the number of flavor, the function $PolyLog[2,z]$ having form, $PolyLog[2,z]=\sum_{k=1}^{\infty} \frac{z^k}{k^2}$ and $z_g$ is the quasi-gluon effective fugacity and $z_q$ is quasi-quark effective fugacity. These distribution functions are isotropic in nature.\\
Here, $Q^2_g , Q^2_q \leq g^2T$, since it acquire the ideal value $g^2T$ asymptotically. The equilibrium distribution functions for quarks and gluons have been extracted from the equation of state (EoS), which in turn will allow a determination of the transport and other bulk properties of the quark gluon-plasma. Simultaneously, the method also yields a quasiparticle description of interacting quarks and gluons. The first EoS is perturbative in the QCD coupling constant and has contributions of $O(g^5)$. The second EoS is an improvement over the first, with contributions up to $O[g^6 ln(1/g)]$; it incorporates the non-perturbative hard thermal contributions. The interaction effects are shown to be captured entirely by the effective chemical potentials for the gluons and the quarks, in both cases. As mentioned earlier, the effective fugacities, $z_g$ and $z_q$ are obtained for EoS1, EoS2 and LEoS. Both $z_g$ and $z_q$ have a very complicated temperature dependence and asymptotically reach to the ideal value unity. In our present analysis, we have used the quasi-particle Debye mass $m_D^{QP}$ Eq.(\ref{eq16}) depending upon the temperature and finite quark-chemical potential for different number of flavors.\\

\section{Binding energy and dissociation temperature}
In this section, we have studied the effect of the finite quark-chemical potential on the binding energies and the dissociation temperature of c$\bar{c}$ and b$\bar{b}$ resonances with different flavors number. To reach this end we solve the Schrodinger equation for the complete understanding of the quarkonia in the hot QGP medium. The binding energy of charmonium and bottomonium state at T=0 is defined by the difference of energy between the $m_Q$ (mass of quarkonia) and the bottom/open charm threshold. But distance between the continuum threshold and the peak position is defined the binding energy at finite value of temperature~\cite{AgnesMocsy}. Hence, the solution of Schrodinger equation for the potential given by Eq.(\ref{eq4}) gives the energy eigen values for the ground and excited states of charmonium and bottomonium i.e, $J/\psi$, $\psi^{\prime}$, $\Upsilon$ and $\Upsilon^{\prime}$ as:
\begin{equation}
\label{eq19}
 E_n = -\frac{1}{n^2}\frac{m_Q\sigma^2}{m^4_D}
\end{equation}
It has been observed that the binding energy decreases with the temperature and also with finite quark-chemical potential which is shown in the figures, \ref{fig.3}, \ref{fig.4}, \ref{fig.5} and \ref{fig.6}. The binding energy of charmonium and bottomonium state at particular values of temperature becomes smaller or equal to the value of mean thermal energy; i.e, the state of quarkonia is said to be dissociated at that given value of temperature. The dissociation temperature for the states of charmonium and bottomonium is also discussed in~\cite{P.Sandin, prd2016_vin, prd2018_vin}. The dissociation temperature of $J/\psi$ and $\Upsilon$ with different number of flavors at different values of the finite quark-chemical potential has been obtained by using the condition: $\Gamma$= 2(B.E). The intersecting point between the binding energy curve and the thermal width of a quarkonium states ($J/\psi$ and $\Upsilon$) is taken as dissociation point of such states.\\

\section{Mass Spectra of Quarkonium state}
For calculating the mass spectra of heavy quarkonia the relation is:
\begin{equation}
\label{eq20}
M=2m_{Q} + B.E
\end{equation}
Here, mass spectra is equal to the sum of the exact formula of energy and twice the quark-mass. Now, we substitute the values of $E_{nl}^{n}$ (B.E) in the above equation,
\begin{equation}
\label{eq21}
M=2m_{Q}+\frac{1}{n^2}\frac{m_Q\sigma^2}{m^4_D}
\end{equation}
Where, $m_{Q}$ represents the  the mass of quarkonium state, such as charmonium and bottomonium mass and $n$ is used for the states of quarkonium (n= 1 for ground states and n= 2 for excited states).\\
\begin{table}
\label{table-I}
\centering
\caption{The dissociation temperature ($T_{D}$) of the ${J/\psi}$ and ${\Upsilon}$ (in units of $T_{c}$) for using fugacity parameters of EoS2 at $N_f$= 1 has been calculated for the different values of finite quark-chemical potential.}
{\begin{tabular}{@{}cccc@{}} \toprule
$State$ & $\mu$=300 MeV & $\mu$=325 MeV & $\mu$=350 MeV\\ \colrule
$J/\psi$ & 0.8637 & 0.8583 & 0.8531\\ \colrule
$\Upsilon$ & 0.9328 & 0.9223 & 0.9139\\ \botrule
\end{tabular}}
\end{table}
 
\section{Results and Discussion}
In the present work we have investigated the properties of the charmonium and the bottomonium at finite temperature and quark-chemical potential using quasi particle approach. Figure \ref{fig.1} shows the variation of the potential with distance (r) for different form of the  Debye mass (e.g. Quasi-Particle, Leading order and Lattice parametrized) at finite quark-chemical potential $\mu$= 300 MeV and $T$= 150 MeV with different number of flavours (i.e., $N_{f}$= 2 [in figure \ref{fig.1} (a)] and 3 [in figure \ref{fig.1} (b)]). It has been found that the variation of potential increases with the distance (r). This potential is not similar to the lattice free energy of heavy-quark in the deconfined phase, which is a well known coulomb potential~\cite{H.Satz}, the Cornell potential is solvable by one-dimensional Fourier-transform (FT) in the hot QCD medium and has a similar form that has been used to study the quarkonium properties (which is consider like the color flux tube structure).\\
\begin{table}
\label{table-II}
\centering
\caption{The dissociation temperature ($T_{D}$) of the ${J/\psi}$ and ${\Upsilon}$ (in units of $T_{c}$) for using fugacity parameters of EoS2 at $N_f$= 2 has been calculated for the different values of finite quark-chemical potential.}
{\begin{tabular}{@{}cccc@{}} \toprule
$State$ & $\mu$=300 MeV & $\mu$=325 MeV & $\mu$=350 MeV\\ \colrule
$J/\psi$ & 0.8652  & 0.8596 & 0.8546\\ \colrule
$\Upsilon$ & 0.9353 & 0.9253 & 0.9164\\ \botrule
\end{tabular}}
\end{table}
\begin{figure*} 
    \includegraphics[height=6.5cm,width=7cm]{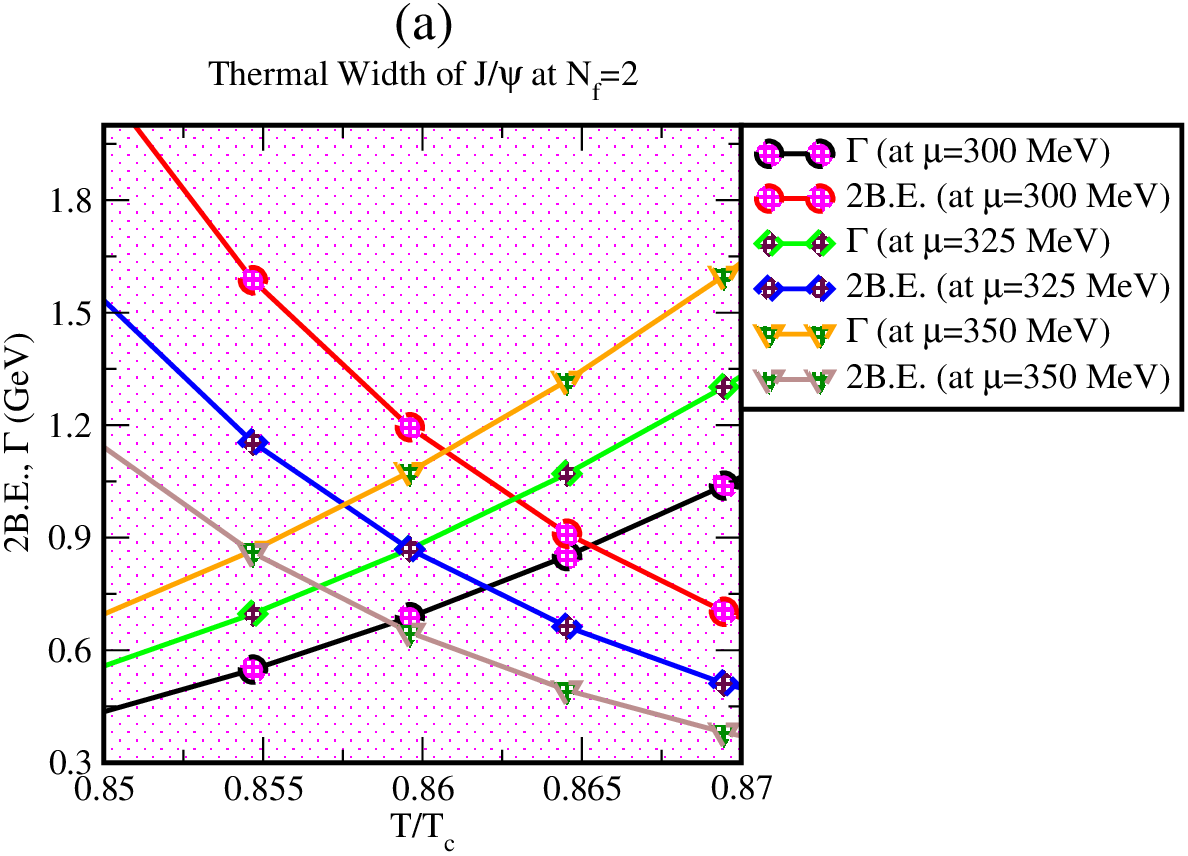}
     \hspace{1cm}
    \includegraphics[height=6.5cm,width=7cm]{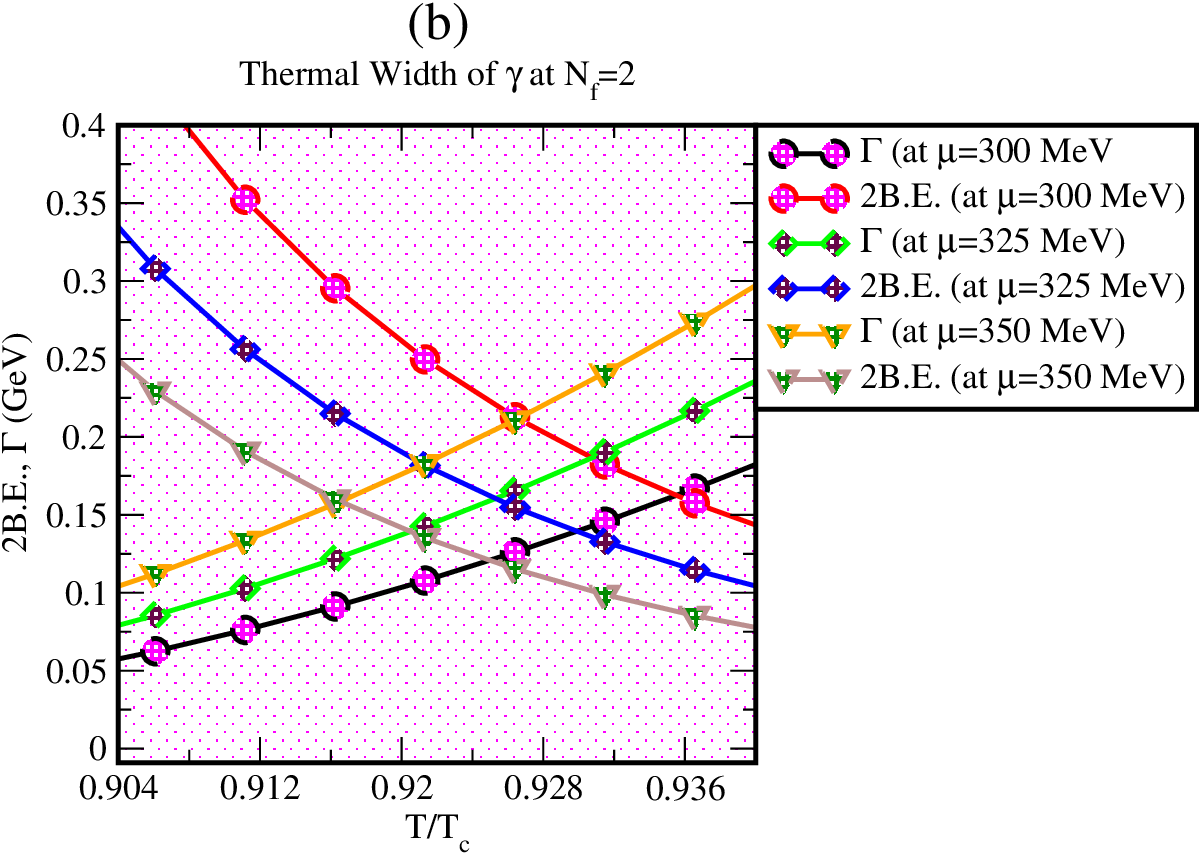}
    \vspace{5mm}
\caption{Shows the real and imaginary binding energies of the $J/\psi$ and $\Upsilon$ for different finite quark-chemical potential at $N_f$= $2$ in the figures 9 (a) and 9 (b).}
\label{fig.9}
\vspace{2cm} 
\end{figure*}
\begin{figure*}  
\vspace{2cm}
    \includegraphics[height=6.5cm,width=7cm]{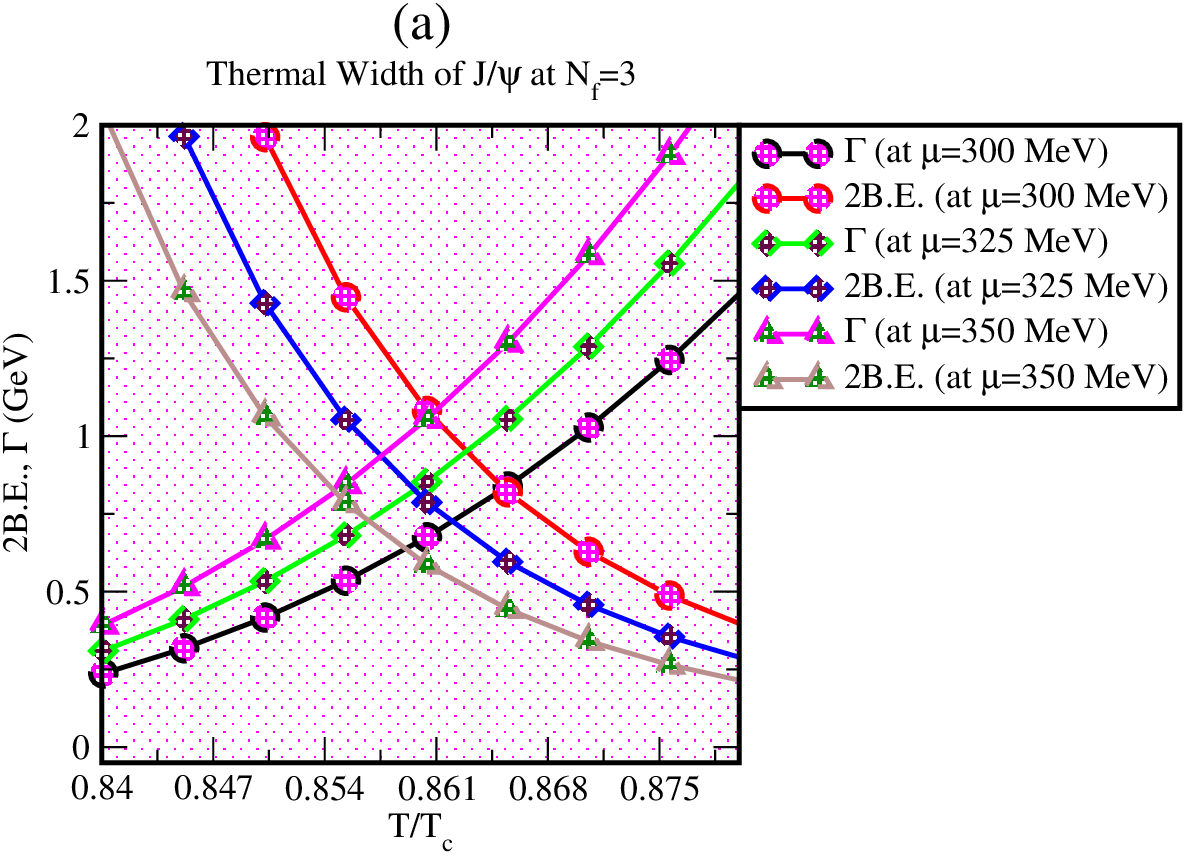}
     \hspace{1cm}
    \includegraphics[height=6.5cm,width=7cm]{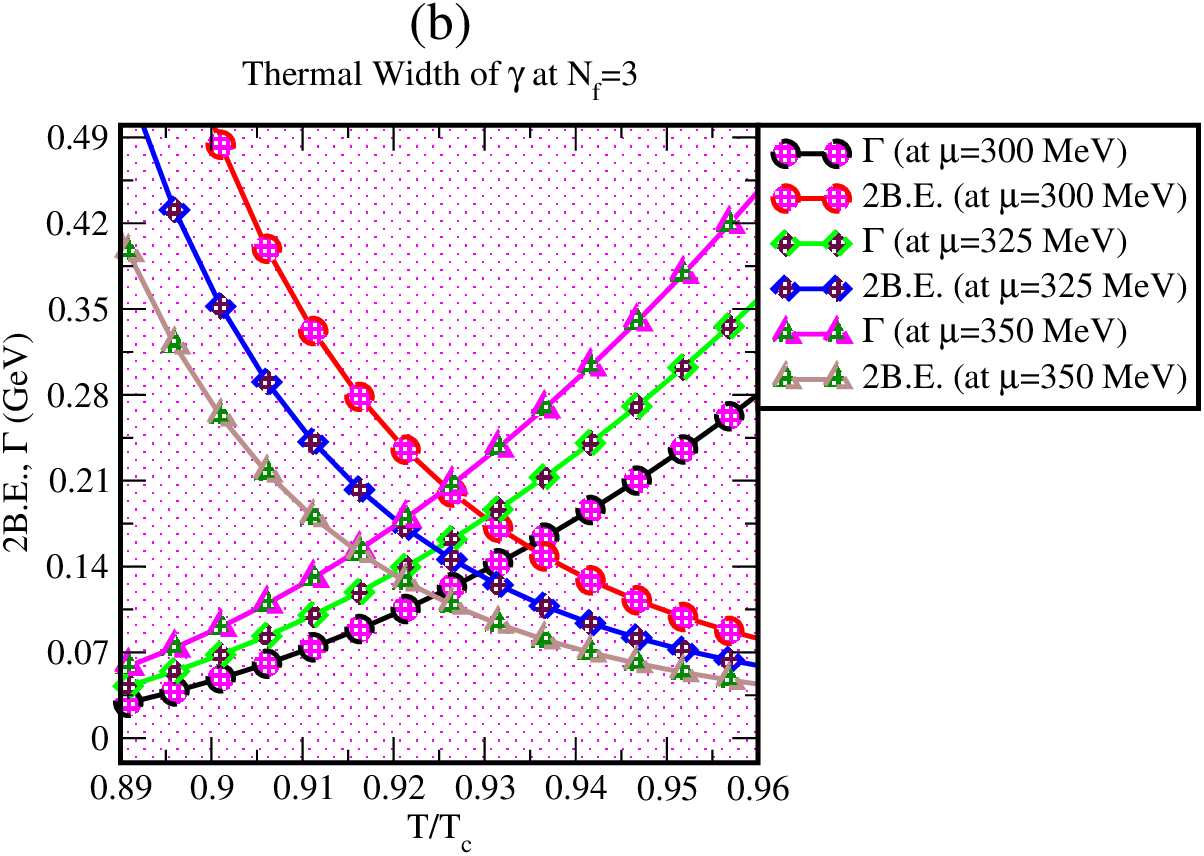}
     \vspace{5mm}
\caption{Shows the real and imaginary binding energies of the $J/\psi$ and $\Upsilon$ for different finite quark-chemical potential at $N_f$= $3$ in the figures 10 (a) and 10 (b)}
\label{fig.10}
\vspace{5mm}  
\end{figure*}
The variation of quasi-particle Debye mass with finite quark-chemical potential at different values of temperature (i.e, 150, 160 and 170 MeV) is shown in figure \ref{fig.2} (a), and the variation of quasi-particle Debye mass with temperature at different values of finite quark-chemical potential (i.e, 150, 200 and 250 MeV) is shown in the figure \ref{fig.2} (b) respectively. The Debye screening mass at baryon density and temperature has been studied by the lattice Taylor expansion method~\cite{M.Doring}. If we increase the value of finite quark-chemical potential, Debye mass increases for different temperatures is shown in the figure \ref{fig.2} (a). But in the figure \ref{fig.2} (b), Debye mass increases with the temperature for different quark-chemical potential. The variation of the binding energy for the $J/\psi$ and $\Upsilon$ with the temperature for different number of the flavors $N_f$= 2 and 3 has been shown in figure \ref{fig.3} (a) and in figure \ref{fig.3} (b) respectively. When we increase the value of number of flavors in the figure \ref{fig.3}, the variation of binding energy decreases at $\mu$= 100 MeV.\\
\begin{table}
\label{table-III}
\centering
\caption{The dissociation temperature ($T_{D}$) of the ${J/\psi}$ and ${\Upsilon}$ (in units of $T_{c}$) for using fugacity parameters of EoS2 at $N_f$= $3$ has been calculated for the different values of finite quark-chemical potential.}
{\begin{tabular}{@{}cccc@{}} \toprule
$State$ & $\mu$=300 MeV & $\mu$=325 MeV & $\mu$=350 MeV\\ \colrule
$J/\psi$ & 0.8651 & 0.8596 & 0.8545\\ \colrule
$\Upsilon$ & 0.9348 & 0.9245 & 0.9160\\ \botrule
\end{tabular}}
\end{table}
Figures \ref{fig.4} and \ref{fig.5} shows the variation of binding energy of $J/\psi$ and $\psi^{\prime}$ in the figure \ref{fig.4} (a) and \ref{fig.5} (a), $\Upsilon$ and $\Upsilon^{\prime}$ in the figure \ref{fig.4} (b) and \ref{fig.5} (b) respectively, with the temperature at $N_f$= 3 for different values of the finite quark-chemical potential ($\mu$= 100, 110 and 120 MeV). From these figures, we notice that the binding energy $J/\psi$, $\psi^{\prime}$, $\Upsilon$ and $\Upsilon^{\prime}$ decreases with the temperature as we increase the value of finite quark-chemical potential. So, it become an interesting fact about the rate of exponential decay of binding energy. This behavior of binding energy could be understood by the strongerness of Debye screening with increasing values of the finite quark-chemical potential $(\mu)$ and also the strength of inter-quark potential, which is weaker as compared to the zero finite quark-chemical potential $\mu$= 0. The binding energy at finite value of temperature and  quark-chemical potential gives information about the dissociation of the quarkonium states (charmonium and bottomonium states). Since, it is known that the binding energy is directly proportional to the mass of quarkonia so, if the value of quarkonia mass increases, in other words we turns towards higher masses i.e. from  mass of $J/\psi$ (1.5 GeV) to mass of $\Upsilon$ (4.5 GeV), we notice that the binding energy increases. The mass spectra of the quarkonia states in the presence of finite quark-chemical potential has been also studied and its variation with the temperature has been shown in the figures \ref{fig.6}. It has been observed that the mass spectra of the $J/\psi$ in the figure \ref{fig.6} (a) and $\Upsilon$ in the figure \ref{fig.6} (b) increases as the mass of the quarkonia increases. We have also made a comparison of the masses of the $J/\psi$ and $\Upsilon$ with the experimental data~\cite{Tanabashi} and this has been given in the Table-IV. The variation of the thermal width of $J/\psi$ in the figure \ref{fig.7} (a) and $\Upsilon$ in the figure \ref{fig.7} (b) with the temperature for different values of the finite quark-chemical potential (i.e, 200, 250 and 300 MeV) has been shown. It has been seen that the thermal width of the quarkonium states with the finite quark-chemical potential increases. The real and the imaginary part of the binding energies of the $J/\psi$ in the figures \ref{fig.8} (a), \ref{fig.9} (a) and \ref{fig.10} (a) and $\Upsilon$ in the figures \ref{fig.8} (b), \ref{fig.9} (b) and \ref{fig.10} (b) with the $T/T_{c}$ for $N_f$= 1, 2 and 3 at $\mu$= 300, 325 and 350 MeV has been shown respectively.
\begin{table}
\label{table-IV}
\centering
\caption{Comparison of the mass spectra for $J/\psi$ and $\Upsilon$ obtained in the present work with the theoretical and experimental data.}
{\begin{tabular}{@{}cccc@{}} \toprule
$State$ & present work (GeV) & Exp.mass\cite{Tanabashi} (GeV) & Error (GeV)\\ \colrule
$J/\psi$ & 3.1  & 3.096 & 0.00129\\ \colrule
$\Upsilon$ & 9.5  & 9.460 & 0.00421\\ \botrule
\end{tabular}}  
\end{table}
From these figures we have examined the dissociation temperature (By the intersection point of the thermal width and two times of the binding energy of the quarkonia) of the $J/\psi$ and $\Upsilon$ for different flavors $N_f$= 1, 2 and 3 and the obtained values of the dissociation temperatures, given in the Table-I, II and III respectively. The dissociation temperature of the $J/\psi$ and $\Upsilon$ at $N_f$ = 1 for the finite quark-chemical potential $\mu$= 300, 325 and 350 MeV are found 0.8637, 0.8583, 0.8531 and 0.9328, 0.9223, 0.9139 respectively (in terms of $T_c$). Similarly at $N_f$= 2 the dissociation temperature of the $J/\psi$ and $\Upsilon$ for different finite quark-chemical potential $\mu$= 300, 325 and 350 MeV  are 0.8652, 0.8596, 0.8546 and 0.9353, 0.9253, 0.9164 respectively (in unit of $T_c$). Finally, the dissociation point of the $J/\psi$ and $\Upsilon$ for different values of the finite quark-chemical potential $\mu$= 300, 325 and 350 MeV was found as 0.8651, 0.8596, 0.8545 and 0.9348, 0.9245, 0.9160 respectively (in unit of $T_c$) when the number of flavor was taken as $N_f$= 3.\\

\section{Conclusion and future outcomes}
The current work have determined the quarkonium dissociation behavior in the hot and dense QGP medium, in the presence of the finite quark-chemical potential. It has been observed that the Cornell potential with distance 'r', increases for the different Debye screening. The behavior of binding energy with temperature has also seen by the earlier studies~\cite{prd2016_vin} and with anisotropic parameter at zero finite quark-chemical potential in~\cite{prd2018_vin}. Here we consider the finite values of the quark-chemical potential. It is noticed that the binding energy for the different quarkonium states decreases with the temperature as we increase the value of finite quark-chemical potential. On the other hand, the dissociation temperature of the quarkonium states  decreases with increasing the values of finite quark-chemical potential. This can be seen from the Table-I, II and III for number of flavor $N_f$= 1, 2 and 3 respectively. The maximum error in the mass spectra of the $J/\psi$ and $\Upsilon$ deduced by using Chi square function is $0.00129$ and $0.00421$ (in GeV) respectively after compared with the experimental values.\\ 
The present work might be helpful in exploring the studies of the compact objects like neutron stars. Since the Compressed Baryonic Matter (CBM) experiment at FAIR is exploring the quark gluon plasma at higher baryon densities, so such type of theoretical studies may contribute to the physics of compact bodies with high baryon densities.\\

\section*{Data Availability}
This is a theoritical work and all previous results are listed in the references.\\

\section*{Conflicts of Interest}
The authors declare no competing interest.\\

\section{Acknowledgments}
One of the author, V.K.Agotiya acknowledge the Science and Engineering Research Board (SERB) Project No. EEQ/2018/000181 New Delhi for the research support in basic sciences.\\

\end{document}